# High-Mobility and High-Reliability Top-Gate Oxide Semiconductor Transistors by Oxygen Engineering

Kai Jiang, Zhiyu Lin, Ziheng Wang, Chen Wang, Mengwei Si

***Abstract*—In this work, we investigate the role of oxygen (O) on the performance of top-gate (TG) atomic-layer-deposited (ALD) oxide semiconductor transistors. The results reveal distinct defect characteristics and positive bias temperature instability (PBTI) degradation mechanisms between oxygen-rich (O-rich) and oxygen-deficient (O-poor) devices. It is found that an O-rich device fabrication process followed by O-free annealing can effectively achieve TG indium-rich (In-rich) oxide semiconductor transistors with high mobility, high reliability and high stability in hydrogen environment because O-rich process can suppress oxygen vacancies and their interaction with hydrogen, while O-free annealing plays a critical role in minimizing the formation of O-rich defects such as oxygen dimers (O-O bonds). Consequently, TG In-rich transistors with high mobility, steep subthreshold slope, and high PBTI reliability at high temperature are demonstrated. The understanding of O-rich defects provides a new insight to overcome the mobility-stability trade-off.**

## I. Introduction

Oxide semiconductor (OS) transistors are considered as promising candidates as back-end-of-line (BEOL) compatible devices for monolithic 3D integration and dynamic random-access memory (DRAM) applications [1-7]. Recent advances in bottom-gate (BG) OS transistors have demonstrated significant improvements in device performance and reliability [8-15]. However, the mechanism for threshold voltage shift ($\Delta V_{TH}$) in the positive bias temperature instability (PBTI) test is still not clear. Meanwhile, high-mobility OS have deeper conduction-band minimum (CBM) and smaller activation energy, which implies that they have lower tolerance to defects, leading to the mobility-stability trade-off [16]. To date, several studies have proposed possible defects for donor in OS, such as oxygen vacancies ($V_O$), hydrogen (H), and the interaction between $V_O$ and H [17-23], but there is still no consensus on the exact nature of the defects involved or their specific impact on reliability. In fact, $\Delta V_{TH}$ during PBTI test arises from interplay of various defects and similar $\Delta V_{TH}$ trends may originate from distinct physical process, making it difficult to investigate the intrinsic mechanism.

Manuscript received xxx xx, xxxx. This work was supported by the National Key Research and Development Program of China under Grant 2022YFB3606900, National Natural Science Foundation of China under Grant 62274107, 92264204 and Shanghai Pilot Program for Basic Research-Shanghai Jiao Tong University under Grant 21TQ1400212. *(Corresponding author: Mengwei Si.)*

Kai Jiang, Zhiyu Lin, Ziheng Wang, Chen Wang, Mengwei Si are with State Key Laboratory of Micro and Nano Engineering Science and the School of Information Science and Electronic Engineering, Shanghai Jiao Tong University, Shanghai 200240, China (email: mengwei.si@sjtu.edu.cn).

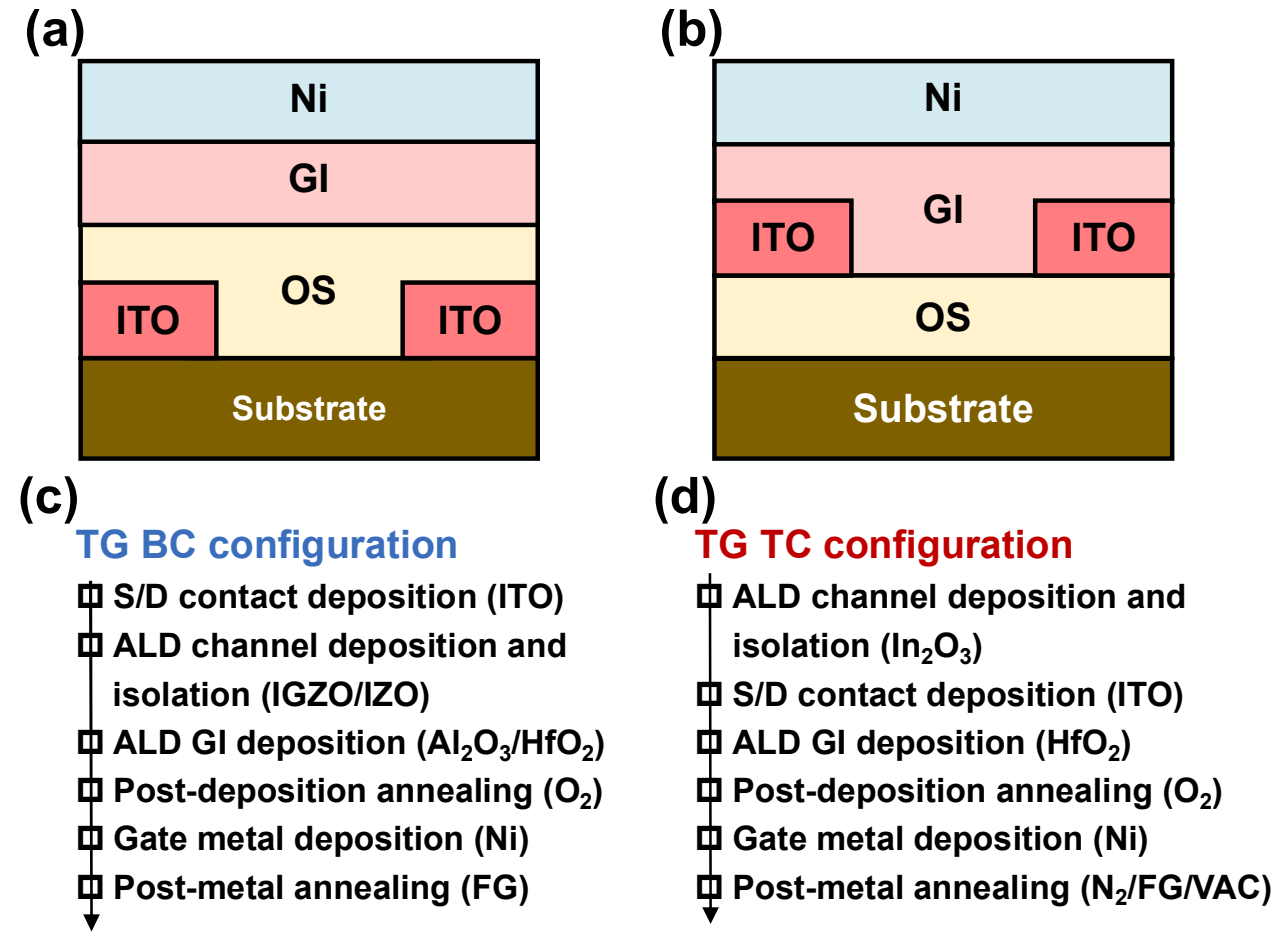

Fig. 1 (a) Schematic diagram of TG (a) BC and (b) TC OS transistors. The device fabrication process flow of (c) TG BC and (d) TG TC devices.

Compared with BG OS transistors, top-gate (TG) OS transistors are more preferred due to their superior integration density and lower parasitic capacitance. For TG structure, the main challenge lies in the generation of $V_O$ during atomic layer deposition (ALD) of gate insulator (GI) on OS channels, caused by metal precursor-induced oxygen scavenging [24-36]. Such an effect is particularly severe for high-mobility indium-rich (In-rich) OS, as In-O bonds exhibit relatively low stability during the GI deposition process [37], [38]. And the interaction between $V_O$ and H results in performance degradation and poor PBTI reliability and stability in H environment. Therefore, although TG OS transistors with relatively low-mobility channels have been reported with high reliability [39-44], the reliability characteristics of TG high-mobility In-rich OS devices remain insufficiently investigated.

In this work, the distinct difference in defect characteristics and PBTI degradation mechanisms between oxygen-rich (O-rich) and oxygen-deficient (O-poor) devices is revealed. It is found both O-rich and O-poor defects lead to similar negative $V_{TH}$ shift during device fabrication, so that confusing results and explanation are commonly reported in literature. We propose and verify that an O-rich device fabrication process followed by oxygen-free (O-free) annealing can simultaneously suppress both O-poor and O-rich defects to achieve high-

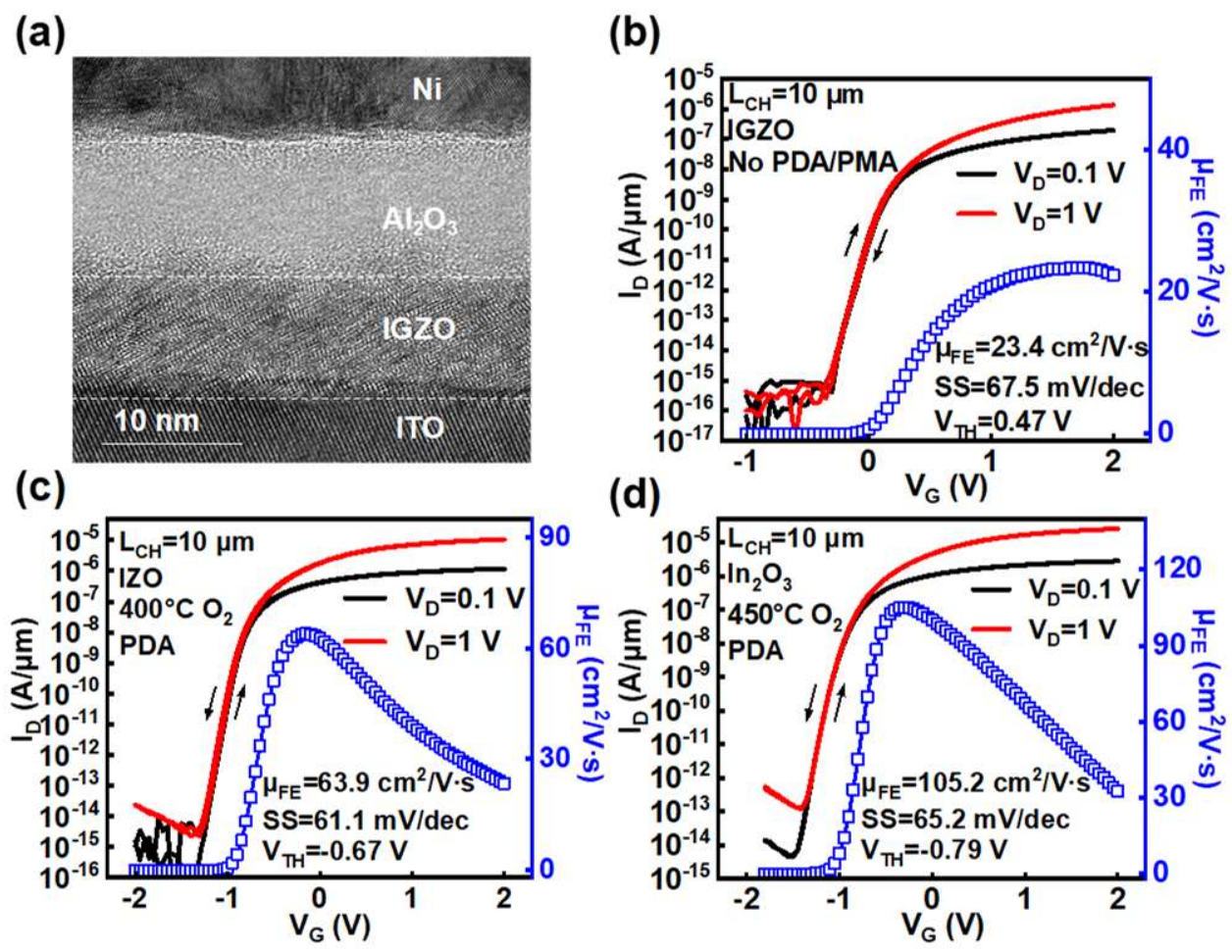

Fig. 2 (a) The cross-sectional TEM image of the S/D region in a TG BC IGZO transistor. $I_D$-$V_G$ and $\mu_{FE}$-$V_G$ characteristics of TG (b) IGZO, (c) IZO, and (d) $In_2O_3$ transistors with $L_{CH}$ of 10 μm. TG In-rich devices show high performance since $O_2$ PDA effectively suppresses the concentration of $V_O$.

mobility TG In-rich OS transistors with enhanced reliability and stability in H environment. The O-rich fabrication process is essential for suppressing $V_O$ and their interaction with H, while the O-free annealing is critical for minimizing the formation of O-rich defects. As a result, we demonstrate TG ALD $In_2O_3$ transistors with high field-effect mobility ($\mu_{FE}$), steep subthreshold slope (SS) and high PBTI reliability at high temperature. The understanding of O-rich defects can provide a new insight to overcome the mobility-stability trade-off.

## II. Experiments

Figs. 1(a) and 1(b) exhibit the schematic diagram of the TG OS transistors with bottom-contact (BC) and top-contact (TC) configurations. The corresponding device fabrication process flow is shown in Figs. 1(c) and 1(d). For BC configuration, source/drain (S/D) electrodes are deposited before channel deposition while S/D electrodes deposition after channel deposition for TC configuration. The device fabrication started with a Si substrate with 90 nm thermally grown $SiO_2$ and 10 nm ALD $Al_2O_3$ as buffer layer. For the BC structure, 80 nm ITO was deposited by sputtering on substrate as S/D electrodes. Then 10 nm IGZO or 8 nm IZO was deposited as OS channel. Channel isolation was achieved through wet etching with hydrochloric acid (HCl). For the TC structure, 4 nm $In_2O_3$ was deposited by ALD as OS channel, followed by channel isolation. Subsequently, 80 nm ITO was deposited by sputtering as S/D electrodes. 10 nm ALD $Al_2O_3$ using trimethylaluminum (TMA) and $H_2O$ as precursors for IGZO channel or 8 nm ALD $HfO_2$ with tetrakis(dimethylamido)hafnium (TDMAHf) and $O_3$ as precursors for IZO and $In_2O_3$ channels was deposited as GI. The growth temperature of both $HfO_2$ and $Al_2O_3$ is 200°C. PDA was conducted in $O_2$ at various temperatures for 10 min. 40 nm Ni was deposited by thermal evaporation as gate metal. Finally, PMA was performed in $N_2$ or vacuum (VAC) or forming gas (FG, 4% $H_2$/96% $N_2$) at various temperatures for 10 min. The cross-sectional transmission electron microscopy (TEM) image in a typical TG BC IGZO transistor is shown in Fig. 2(a).

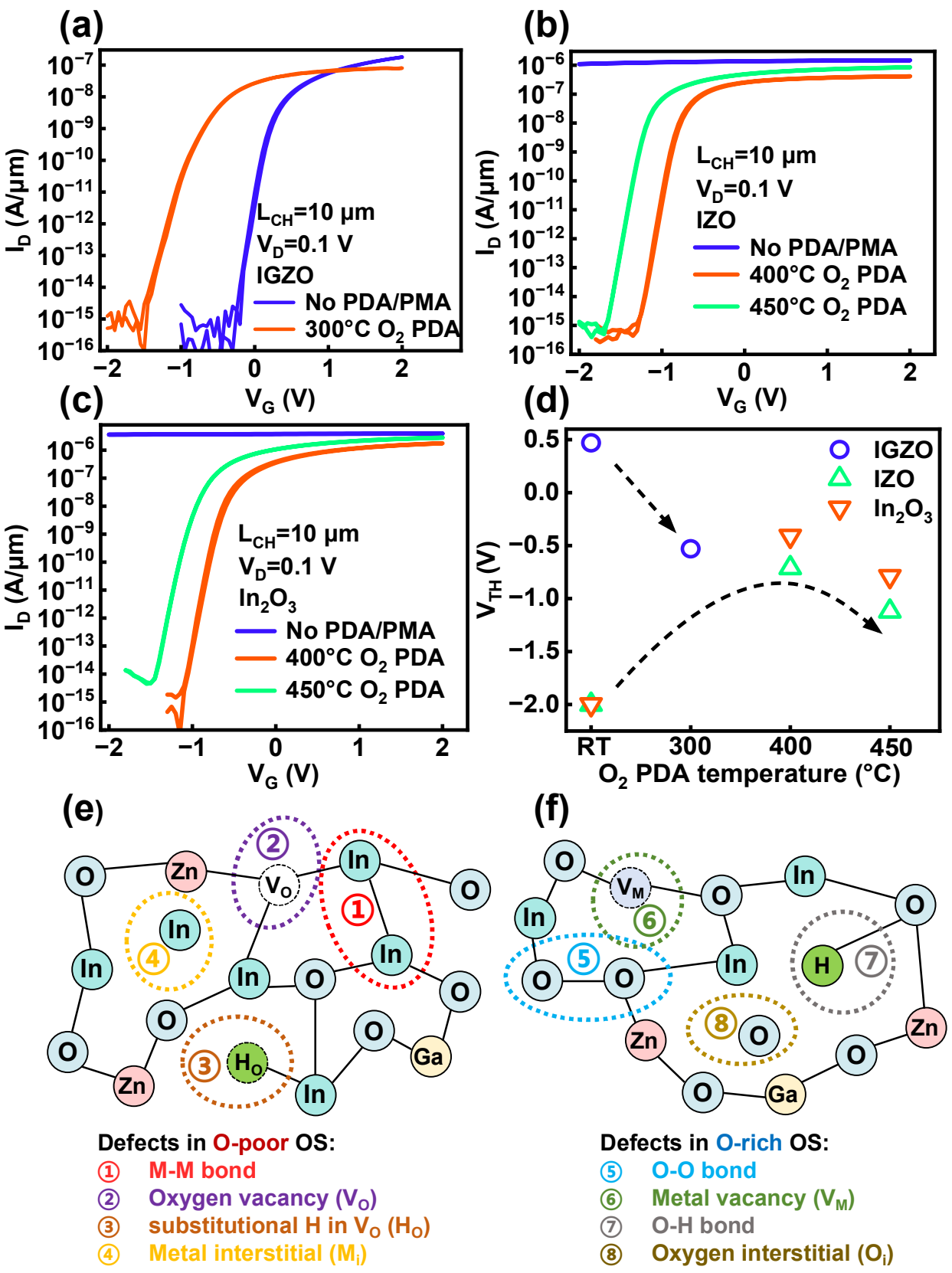

Fig. 3 Transfer characteristics of TG (a) IGZO, (b) IZO, and (c) $In_2O_3$ transistors under different annealing conditions. (d) $V_{TH}$ versus $O_2$ PDA temperature extracted from Figs. 3(a) to 3(c). IZO and $In_2O_3$ devices without annealing cannot be turned off and $V_{TH}$ of -2 V is used here for comparison. The possible defect types in (e) O-poor and (f) O-rich OS films. The dominant defect types differ fundamentally between O-rich and O-poor films.

## III. Results and Discussion

Figs. 2(b) to 2(d) exhibit the $I_D$/$\mu_{FE}$-$V_G$ curves of TG OS transistors with channel length ($L_{CH}$) of 10 μm and different channel materials. TG IGZO transistors without annealing exhibit large on/off ratio. Moreover, although $H_2O$ was used as precursor for $Al_2O_3$ GI, the devices without annealing still exhibit decent SS and on/off ratio. This result indicates only minimal degradation of the IGZO channel during GI deposition, which can be partly explained by the presence of Ga enhances the stability of IGZO, but the role of H here is not clear. In contrast, TG In-rich IZO and $In_2O_3$ devices without annealing cannot be turned off, indicating oxide atoms are taken away from the oxide channel layer due to oxygen scavenging by Hf precursors, resulting in damage to OS channel layer. High-temperature $O_2$ annealing can effectively suppress the concentration of $V_O$ and the TG In-rich devices achieve high $\mu_{FE}$, steep SS, and negligible hysteresis.

Oxygen PDA exerts a profound influence on device performance. Systematic annealing under varying conditions was therefore conducted to investigate the underlying mechanisms. Figs. 3(a) to 3(c) exhibit the transfer characteristics of TG OS transistors with different OS channels under different annealing conditions. For the TG IGZO transistor, $O_2$ PDA induces a substantial negative $\Delta V_{TH}$ and degradation of SS. For both IZO and $In_2O_3$ channels, devices without annealing cannot be turned off, originating from $V_O$ generation during GI deposition, where Hf precursors scavenge oxygen from OS channel layer, as previously reported [24], [37], [38]. With 400°C $O_2$ PDA, the devices exhibit a significant more positive $V_{TH}$, which is attributed to the effective suppression of $V_O$. However, as the $O_2$ PDA temperature rises to 450°C, an abnormal negative $\Delta V_{TH}$ is also observed, indicating the presence of donor-type O-rich defects, with their generation mechanism attributed to high-temperature $O_2$ annealing. Fig. 3(d) illustrates $V_{TH}$ of the devices with different OS channels as a function of $O_2$ PDA temperature, clearly demonstrating an initial positive $\Delta V_{TH}$ followed by a negative $\Delta V_{TH}$ with increasing $O_2$ PDA temperature for both TG IZO and $In_2O_3$ transistors, indicating a transition from O-poor to O-rich state in the OS channels. For the IGZO channel, the device without annealing is not in O-poor state since the presence of Ga enhances the stability of IGZO, while 300°C $O_2$ PDA introduces excess O-rich defects, resulting in a negative $\Delta V_{TH}$.

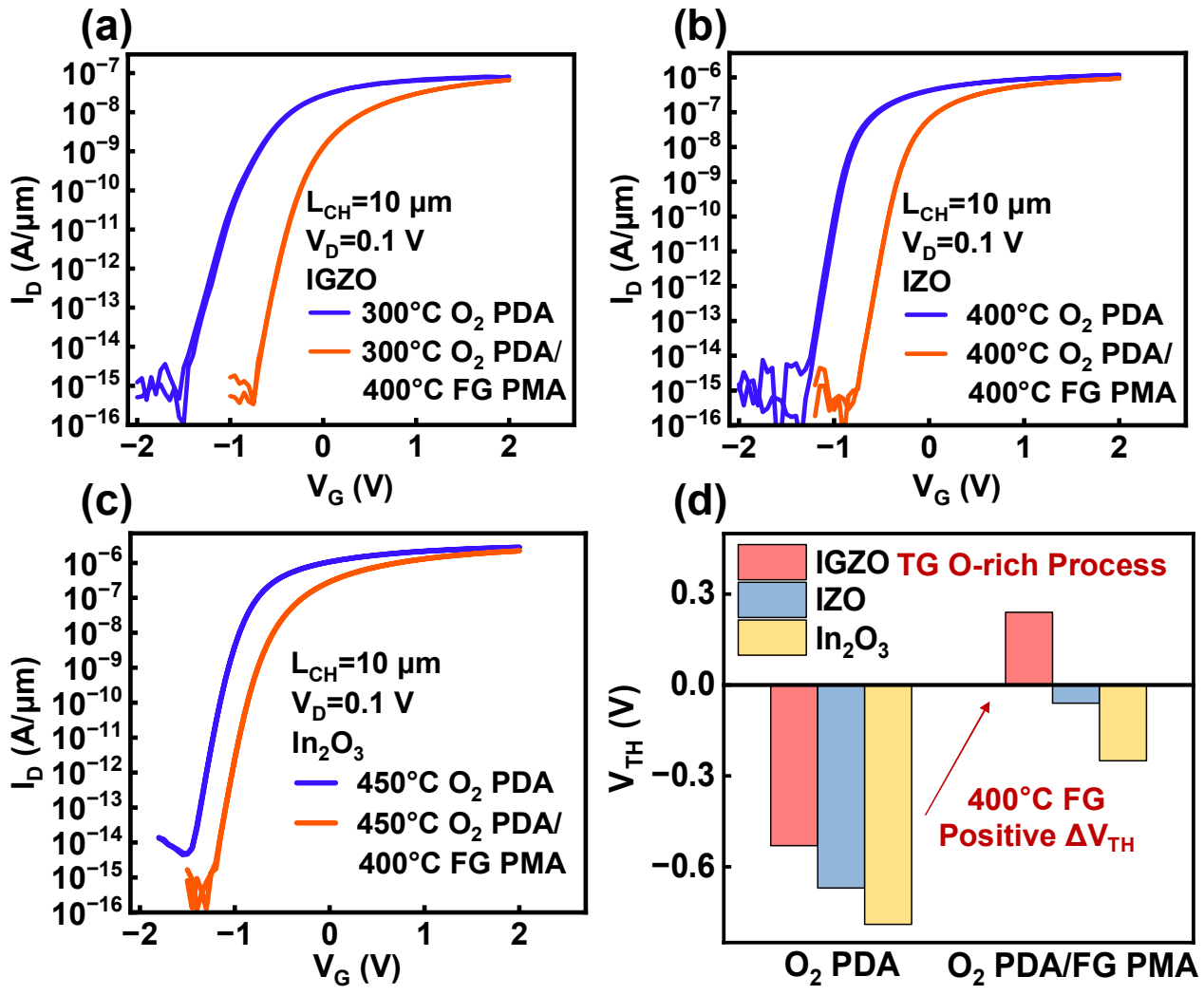

Fig. 4 Transfer characteristics of TG (a) IGZO, (b) IZO, and (c) $In_2O_3$ devices under different annealing conditions. (d) The impact of FG PMA on $V_{TH}$ of TG O-rich devices extracted from Figs. 4(a) to 4(c). All devices exhibit more positive $V_{TH}$ after additional FG PMA.

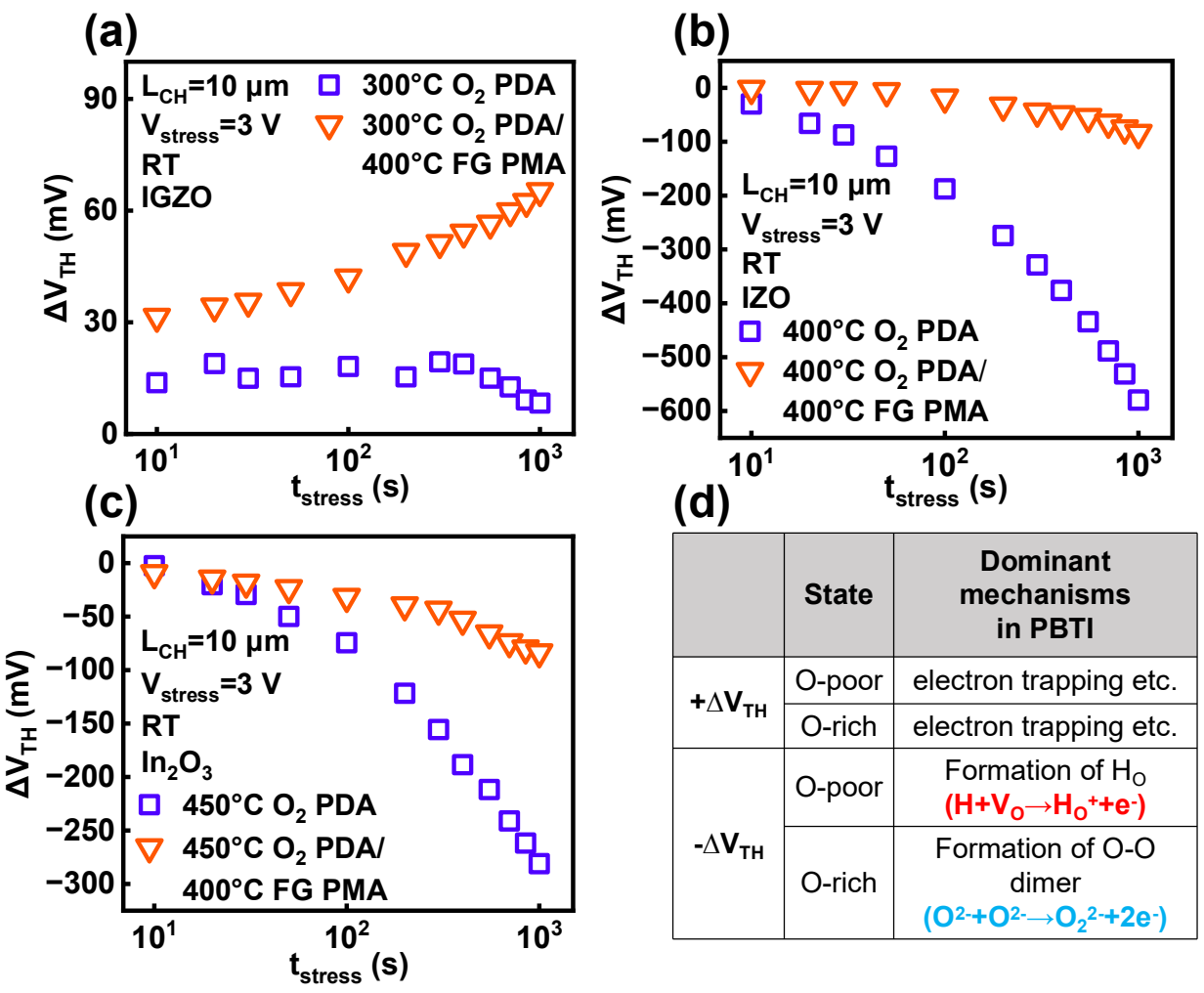

Fig. 5 $\Delta V_{TH}$ versus $t_{stress}$ characteristics at RT of TG (a) IGZO, (b) IZO, and (c) $In_2O_3$ transistors under different annealing conditions. (d) Summary of PBTI degradation mechanisms in OS transistors and the dominant negative $\Delta V_{TH}$ mechanisms differ significantly between O-rich and O-poor devices owing to different defect types.

Based on the above experimental results, the possible defect types in the OS films are summarized in Figs. 3(e) and 3(f). The dominant defect types differ fundamentally between O-rich and O-poor films. Defects such as O-O bond, metal vacancy ($V_M$), O-H bond, metal interstitial ($M_i$), etc. may exist in O-rich film, while defects such as M-M bond, $V_O$, M-H bond (substitutional H in $V_O$, $H_O$), oxygen interstitial ($O_i$), etc. may exist in O-poor film. However, O-rich and O-poor devices may behave similarly even though the distinct defects exist in the OS films. It has been understood that H could react with $V_O$ (O-poor defects), leading to negative $\Delta V_{TH}$ [19-22], [37]. High-temperature $O_2$ annealing could suppress O-poor defects while inducing O-rich defects, which also results in negative $\Delta V_{TH}$ through distinct mechanisms. Therefore, both O-poor and O-rich defects can exhibit donor-like behavior in the OS films, making it difficult to distinguish them.

To verify that the TG OS transistors with $O_2$ PDA are in O-rich state and O-rich defects can exhibit donor-like behavior, devices with $O_2$ PDA were annealed in FG. It has been understood that H could react with $V_O$ (O-poor defects) leading to negative $\Delta V_{TH}$ and devices typically show negative $\Delta V_{TH}$ after FG annealing. Figs. 4(a) to 4(c) present the transfer characteristics of TG OS transistors with FG PMA after $O_2$ PDA. All devices exhibit more positive $V_{TH}$ after additional FG PMA, indicating the interaction between H and $V_O$ has been effectively suppressed since the concentration of $V_O$ is reduced by $O_2$ PDA and the dominant defects in the device are not O-poor defects. Therefore, by O-rich fabrication process, stability in H environment at high temperature can be improved significantly without $V_{TH}$ degradation. Fig. 4(d) summarizes the impact of FG PMA on $V_{TH}$ of TG O-rich devices. The O-rich devices with $O_2$ PDA exhibit more positive $V_{TH}$ after additional FG PMA, indicating that the O-rich defects are donor-like defects and the positive $\Delta V_{TH}$ after FG PMA suggests O-rich defects could be suppressed by O-free annealing.

More positive $V_{TH}$ in the transfer characteristics after FG PMA indicates the devices with $O_2$ PDA are in O-rich state. To further confirm the generation of O-rich defects after high-temperature $O_2$ annealing, PBTI performance was measured. Figs. 5(a) to 5(c) show the PBTI performance of TG OS transistors with different annealing conditions at room temperature (RT). For IZO and $In_2O_3$ channels, devices with only $O_2$ PDA exhibit significant negative $\Delta V_{TH}$. The negative $\Delta V_{TH}$ during PBTI is typically attributed to the interaction between H and $V_O$. However, the devices show high stability in H environment owing to the suppression of $V_O$ by $O_2$ PDA and the negative $\Delta V_{TH}$ here should originate from other mechanisms. Moreover, $\Delta V_{TH}$ becomes more positive with additional FG PMA and the absolute value of $\Delta V_{TH}$ reduces significantly, confirming that the negative $\Delta V_{TH}$ observed in devices with only $O_2$ PDA does not originate from the interaction between $V_O$ and H. FG PMA after $O_2$ PDA leads to more positive $V_{TH}$ in the transfer characteristics and enhanced PBTI reliability. The enhancement of PBTI reliability by FG PMA provides compelling evidence that the negative $\Delta V_{TH}$ during PBTI test in devices with $O_2$ PDA originates from O-

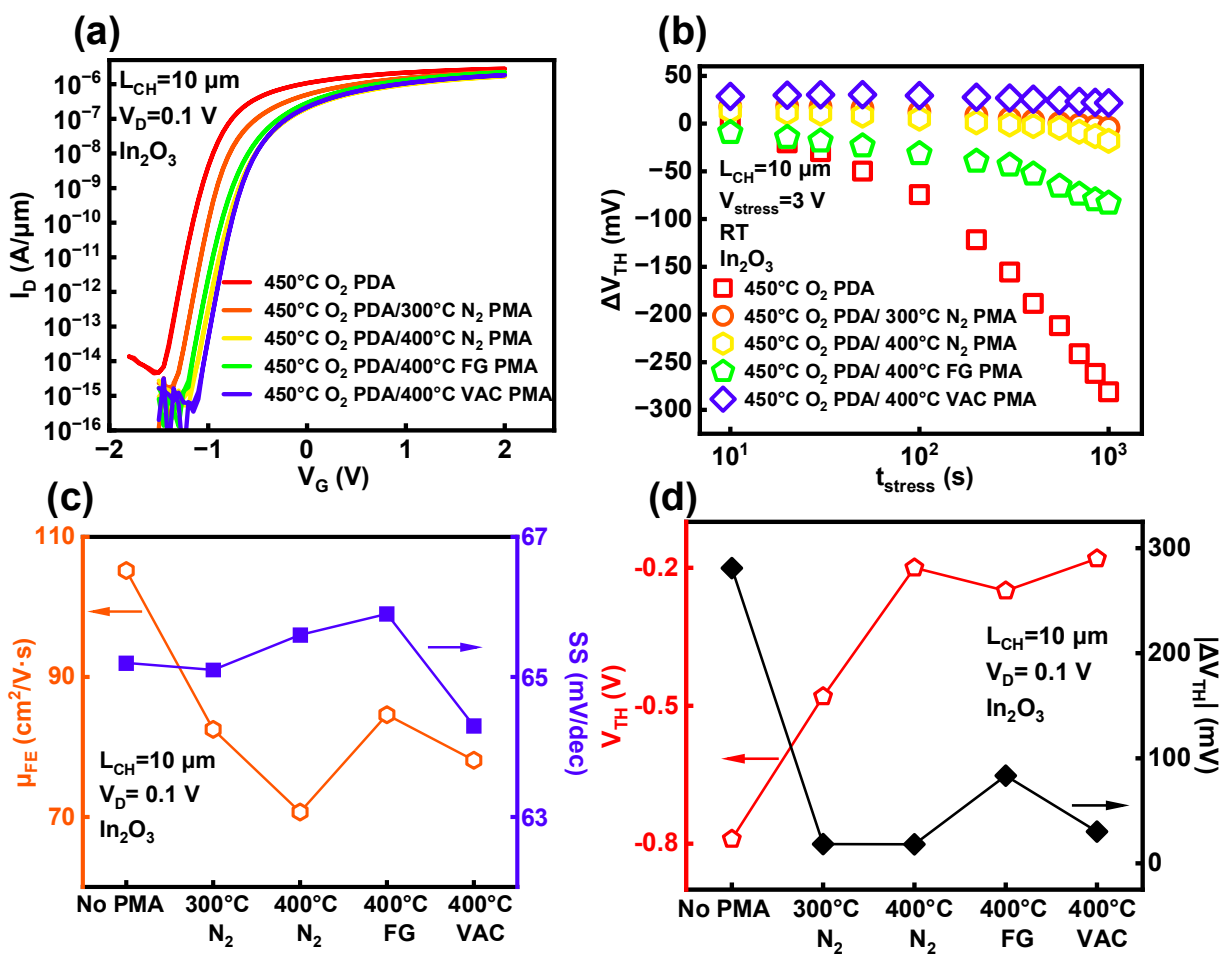

Fig. 6 (a) Transfer characteristics of TG $In_2O_3$ transistors with $L_{CH}$ of 10 µm under different PMA conditions. (b) $\Delta V_{TH}$ versus $t_{stress}$ characteristics at RT of TG $In_2O_3$ transistors with $L_{CH}$ of 10 µm under different annealing conditions. (c) $\mu_{FE}$ and SS, (d) $V_{TH}$ and $|\Delta V_{TH}|$ under PBTI at $V_{stress}$=3 V for 1 ks of TG $In_2O_3$ transistors with $O_2$ PDA and different PMA conditions, extracted from Figs. 6(a) and 6(b).

rich defects rather than the interaction between H and $V_O$. For devices with IGZO channel, additional FG PMA also induces more positive $\Delta V_{TH}$ since the suppression of O-rich defects reduces negative $\Delta V_{TH}$ component. Fig. 5(d) summarizes the possible dominant PBTI degradation mechanisms in OS transistors. The dominant negative $\Delta V_{TH}$ mechanisms during PBTI stress differ significantly between O-rich and O-poor devices due to different defect types.

FG PMA after $O_2$ PDA leads to more positive $V_{TH}$ in the transfer characteristics and enhanced PBTI reliability by suppressing O-rich defects. However, a small amount of $V_O$ may still exist in O-rich devices and the interaction between H and $V_O$ will degrade the device performance. Therefore, PMA under different atmospheres, including $N_2$, FG, and VAC, are performed. The effects of PMA under different atmospheres on TG $In_2O_3$ transistors are shown in Fig. 6(a). PMA under all O-free atmospheres induces a positive $\Delta V_{TH}$, with the magnitude of $\Delta V_{TH}$ increasing with PMA temperature, indicating O-rich defects exhibit donor-like behavior and thermal treatment can effectively suppress O-rich defects. Fig. 6(b) compares PBTI performances of TG $In_2O_3$ transistors under different PMA conditions. All O-free PMA enhance reliability by reducing the magnitude of negative $\Delta V_{TH}$, confirming O-rich defects also induce negative $\Delta V_{TH}$ in PBTI performance. Devices with FG PMA exhibit more negative $V_{TH}$ in the transfer characteristics and $\Delta V_{TH}$ during PBTI stress compared to devices with VAC PMA or $N_2$ PMA, suggesting the existence of a small amount of $V_O$ even in O-rich films, which will react with H after FG PMA. Figs. 6(c) and 6(d) present $\mu_{FE}$, SS, $V_{TH}$ and $|\Delta V_{TH}|$ under PBTI measurement at $V_{stress}$=3 V for 1 ks of the TG $In_2O_3$ transistors under different PMA conditions. Devices with all O-free PMA show more positive $V_{TH}$ in the transfer characteristics and enhanced reliability. By further eliminating the influence of H, devices with $N_2$ or VAC PMA demonstrate high $\mu_{FE}$, steep SS, high reliability and relatively positive $V_{TH}$. The high performance is achieved by $O_2$ PDA and O-free PMA, which suppress both O-poor and O-rich defects.

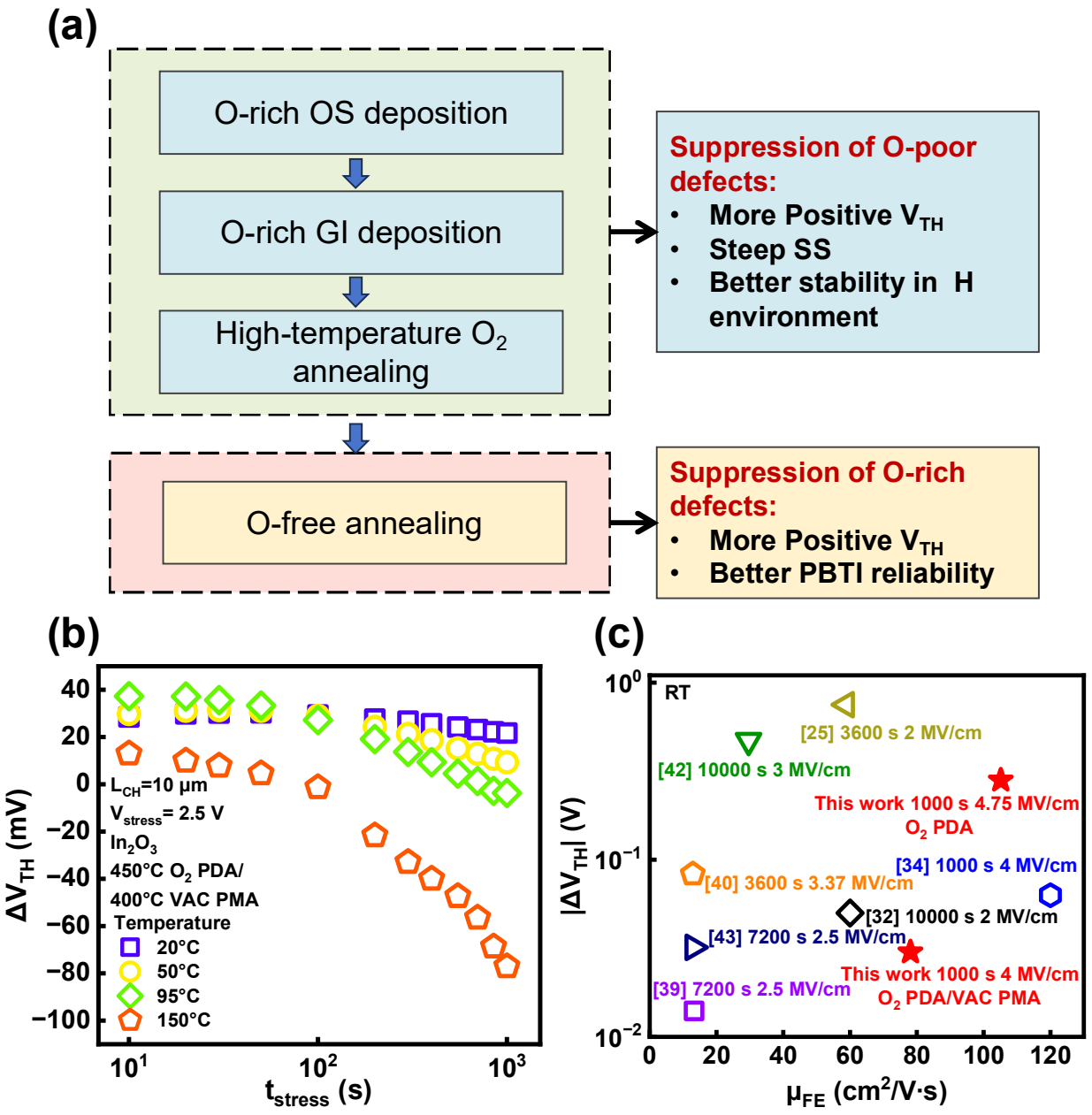

Fig. 7 (a) Strategies for achieving high-mobility and high-stability TG OS transistors. (b) $\Delta V_{TH}$ versus $t_{stress}$ characteristics of TG $In_2O_3$ transistors with $O_2$ PDA and VAC PMA measured at different temperatures. (c) Benchmark of PBTI reliability at RT versus $\mu_{FE}$ of TG OS transistors.

Fig. 7(a) summarizes the effective strategy for enhancing the reliability of high-mobility TG In-rich OS transistors. The O-rich device fabrication process followed by $O_2$ annealing could suppress O-poor defects, leading to more positive $V_{TH}$, steep SS, and better stability in H environment. However, O-rich defects will be introduced during the suppression of O-poor defects, resulting in more negative $V_{TH}$ and poor PBTI reliability. Therefore, O-free annealing is needed to suppress O-rich defects, resulting in more positive $V_{TH}$ and better PBTI reliability. The suppression of both O-poor and O-rich defects improve the device performance, overcoming the mobility-stability trade-off. Fig. 7(b) presents PBTI performance of TG $In_2O_3$ transistors with 450°C $O_2$ PDA and 400°C VAC PMA measured at different temperatures. The high-mobility In-rich device maintains high reliability even at elevated temperatures, due to both O-poor and O-rich defects are suppressed. Fig. 7(c) benchmarks the PBTI reliability of TG OS transistors measured at RT. By further suppress O-rich defects, the devices demonstrate high $\mu_{FE}$ of 78 $cm^2/V \cdot s$ and a low $|\Delta V_{TH}|$ of 30 mV at RT, $V_{stress}$ of 3 V, and $t_{stress}$ of 1 ks. The TG $In_2O_3$ transistors with $O_2$ PDA and O-free PMA in this work exhibit both high reliability and high $\mu_{FE}$, demonstrating the effectiveness of the proposed method, O-rich device fabrication processes followed by O-free annealing, for optimizing the performance of high-mobility TG In-rich OS transistors to overcome the mobility-stability trade-off.

## IV. Conclusion

In conclusion, we demonstrate TG ALD In-rich OS transistors with high $\mu_{FE}$, steep SS, high reliability and high stability in H environment. The performance enhancement is achieved by O-rich device fabrication processes followed by O-free PMA, resulting in the suppression of both O-poor and O-rich defects. The distinct difference in defect characteristics and PBTI degradation mechanisms between O-rich and O-poor devices is clarified. The understanding of different defect types provides a new insight to overcome the mobility-stability trade-off.

## ACKNOWLEDGMENT

The authors also gratefully acknowledge the support from Huawei corp. for IGZO deposition.